\documentstyle[12pt,epsf]{article}

\date{}
\begin{document}
\title{Meson Exchange and Pion Rescattering Contributions to the 
Cross Section for $pp\rightarrow pp\pi^0$}
\vspace{1.5cm}
\author{U. van Kolck$^1$, G.A. Miller$^1$, and D.O. Riska$^2$}
\maketitle
\centerline{$^1$ \it Department of Physics, University of Washington, 
Box 351560,} 

\centerline{\it Seattle, WA 98195-1560, USA}

\centerline{$^2$ \it Department of Physics, 00014 University of
Helsinki, Finland}

\vspace{1cm}

\centerline{\bf Abstract}
\vspace{0.5cm}
Short range meson-exchange mechanisms such as $\rho\to\omega\pi^0$ 
contribute significantly to the amplitude 
for $pp\rightarrow pp\pi^0$ near threshold in addition to pion
rescattering. The uncertainty of the  latter contribution,
which depends on poorly determined parameters, is estimated.
Allowing for this uncertainty and taking the short range meson exchange 
contributions into account makes it
possible to reproduce the measured cross section.

\vspace{4cm}

\hfill{DOE/ER/40427-15-N96} 

\newpage
The reaction $pp\rightarrow pp\pi^0$ near threshold is peculiarly
intriguing because the relatively large value of the momentum transfer
($\approx 360$ MeV/$c$) implies a very small 
impulse approximation term (Fig.~1a). Furthermore, the pion
rescattering contribution (Fig.~1b) is strongly suppressed because 
the dominant low energy pion-nucleon ($\pi N$)
scattering amplitude is the isovector 
Weinberg-Tomozawa term, which cannot contribute to this reaction
\cite{KoRe}. The isoscalar pion rescattering
term can be evaluated using chiral perturbation 
theory (ChPT), which relates this 
contribution to the off-shell $\pi N$
scattering amplitude. Previous calculations \cite{CoFrMiKo,Pa} 
based on a $O(q^3)$ determination of this amplitude {}from 
the isospin-even scattering length and
the sigma-term \cite{BeKaMe1} found that
the rescattering term leads to a significant  
destructive interference with the impulse approximation term.
Part of this destructive interference 
arises {}from intermediate $\Delta(1232)$'s 
(Fig.~1c) \cite{CoFrMiKo}.
The net result is 
a near cancellation between the pion rescattering contribution 
based on the chiral Lagrangian
and the amplitude for single-nucleon 
pion production (this cancellation does not appear in
off-shell extrapolations unconstrained by ChPT \cite{other}). 
As a consequence, the empirical cross section \cite{Me,Mey,Bo} has
to be understood in terms of non-pionic 
two-body mechanisms
of shorter range than that associated with pion rescattering.\\

That shorter-ranged mechanisms should play a significant role in
the reaction $pp\rightarrow pp\pi^0$ was
already implied by the demonstration that the single-nucleon and the
pion-exchange amplitude as conventionally calculated with the (now ruled
out) on-shell approximation for the $\pi N$ scattering amplitude
predict a cross section smaller than the measured one by 
about a factor of 5 \cite{MiSa,Ni}. 
Subsequently it was found that the effective scalar
and vector exchange components of the nucleon-nucleon ($NN$) interaction
imply relativistic two-nucleon operators of short range (Fig.~1d) that
contribute significantly to the cross section 
near threshold \cite{LeRi,HoMeGr}. The same 
mechanisms, through PCAC, had earlier been shown to provide a natural
explanation \cite{KiRiTs,To} of the observed large
enhancement in heavy nuclei of the axial charge of the nucleon
\cite{Wa,War}. The cancellation between the single-nucleon 
and pion-exchange mechanisms shows, however, that the
short-range operators implied by the scalar and vector components of
the $NN$ interaction cannot explain all of
the measured cross section for $pp\rightarrow pp\pi^0$.\\

We consider here 
meson exchange mechanisms in which the
pion is produced at a transition vertex that involves two different
heavy mesons (Fig.~1e): most relevant are 
$\rho\omega$, $\pi ' \sigma$ and $\eta a_0$ exchanges. 
We also consider the effect of 
the intermediate $N(1440)$ resonance that can be
excited by exchanged $\sigma$ and $\omega$ mesons (Fig.~1f).
These contributions have
a sizeable uncertainty range because of the values
(and relative signs) of the various coupling constants are not 
well known.
More dramatic, however, is the theoretical uncertainty 
associated with the pion rescattering contribution (Fig.~1b),
which is estimated by a re-evaluation of
the rescattering contribution
with another set of parameters: a $O(q^2)$ determination {}from $\pi N$ 
sub-threshold parameters \cite{BeKaMe2}. 
In this case, the interference with the one-body term, while 
still destructive, is far less complete. Finally, the 
uncertainty regarding the $NN$
wave functions at small separations is estimated by using
two different realistic, modern $NN$ potentials, 
Argonne V18 \cite{WiStSc} and Reid93 \cite{Fr}.\\

It turns out that the $\rho\omega$ and, to a lesser extent,
$\pi ' \sigma$ and $\eta a_0$ exchange mechanisms  provide
sufficiently large matrix elements for the reaction $pp\rightarrow
pp\pi^0$ so that, when combined with the short range exchange
mechanisms that are associated with the $NN$ interaction,
they can provide an -- at least qualitative -- explanation of the
empirical cross section. \\

Consider first 
the $\rho\omega$ exchange contribution to the two--nucleon pion
production amplitude (Fig.~1e), as derived in Ref. \cite{BeRi}.
The key ingredient in this amplitude is the $\pi\rho\omega$ vertex,
which was considered by Gell-Mann, Sharp and Wagner \cite{GeShWa} in the
$\rho$-dominance model for the decay $\omega\rightarrow \pi\pi\pi$:
\begin{equation}
{\cal L}_{\pi\rho\omega}=-{g_{\pi\rho\omega}\over
m_\omega}\epsilon_{\mu\nu\lambda\delta}\partial^\mu\vec
\rho^{\,\nu}\omega^\lambda\partial^\delta\vec \pi.
\end{equation}
Here $\vec \rho_\nu$ and $\omega_\lambda$ are the Proca field
operators for the $\rho$ and $\omega$ mesons and $\vec \pi$ is the
pion field. The $\pi\rho\omega$ vertex is related to the triangle
anomaly, as illustrated explicitly e.g. in Ref. \cite{MeKaWe}, and
hence it can be viewed as fairly well established. The numerical value
for the coupling constant $g_{\pi\rho\omega}$ is $\simeq -10$
\cite{MeKaWe} (the phase is fixed as in Ref. \cite{SaWa}).\\

The discussion of the 
$\rho\omega$ exchange pion production amplitude is completed by 
noting  that we employ the usual $\omega N\bar N$ and $\rho N\bar N$ 
interactions with vector and tensor coupling.
For the $\omega$- and $\rho$-nucleon
coupling constants we use the values $g_\rho^2/4\pi=1.2$,
$\kappa_\rho=6.1$ and $g_\omega^2/4\pi=25$, $\kappa_\omega=0$
respectively, where, as usual, $\kappa_{\rho(\omega)}$
represents the relative strength of the tensor coupling. 
These values correspond to the Bonn boson-exchange model
for the $NN$ interaction \cite{MaHoEl} and imply
that the $\rho$ and $\omega$ exchanges are treated as effective meson
exchanges rather than pure meson exchange interactions. 
We also assume that 
$g_\rho g_\omega > 0$, in line with the quark model.
The $\rho N\bar N$ and
$\omega N\bar N$ vertex form factors are denoted $F_\rho(\vec{k})$ and
$F_\omega (\vec{k})$ respectively. Their expressions may be inferred {}from
boson exchange models for the $NN$ interaction. 
\\

The complete $\rho\omega$ exchange amplitude for neutral pion
production near threshold is evaluated to be
\begin{eqnarray}
T^{\rho,\omega} & = & -i{g_\rho g_\omega g_{\pi\rho\omega}\over 4m_N^2}
{\omega_q\over m_\omega}
{F_\rho(\vec{k})F_\omega(\vec{k})
\over (\vec{k}^2+m_\rho^2)(\vec{k}^2+m_\omega^2)} \nonumber\\
 & & [(2+ \kappa_\rho + \kappa_\omega) 
(\vec{k}^2\vec{\Sigma}\cdot\vec{P}
-\vec{k}\cdot\vec{P}\vec{\Sigma}\cdot\vec{k}) \nonumber\\
 & & -2i (1+\kappa_\rho)(1+\kappa_\omega)\vec{k}^2 \vec{\sigma}^{(1)} \times
\vec{\sigma}^{(2)} \cdot \vec{k}]. \label{tpirho}
\end{eqnarray} 
Here we follow the notation of Ref. \cite{CoFrMiKo}: 
$\omega_q^2 = \vec{q}^{\, 2} + m_{\pi}^2$ is the energy of the
(on-shell) pion produced with momentum $\vec{q}$ in the center of mass;
$\vec{p}$ ($\vec{p'}$) is the center of mass momentum of the incoming 
(outgoing) proton labelled ``1'' (those of proton ``2'' are opposite);
$\vec{k}= \vec{p} -\vec{p'}$ is the momentum transferred;
$\vec{P}= \vec{p} +\vec{p'}$;
$\vec{\sigma}^{(i)}$ is the spin of proton $i$; and 
$\vec{\Sigma}= \vec{\sigma}^{(1)} - \vec{\sigma}^{(2)}$.  
In Ref. \cite{BeRi} only the numerically most important
last term in this expression was retained.\\

The $\rho\omega$ exchange operator of Eq.~(2) is insignificant for 
nuclear pion absorption rates in near threshold, but it is of
comparable magnitude to the (small) isospin symmetric pion exchange
contribution to the pion production (or absorption) amplitude. As a
consequence it gives a numerically significant contribution to the
matrix element for the reaction $pp\rightarrow pp\pi^0$.\\

In view of the dominant $\pi\eta$ decay mode of the $a_0(980)$ meson
and the important role the effective isospin 1 scalar exchange
mechanism plays in the $NN$ interaction, the $\eta a_0$
exchange mechanism (Fig. 1e) could be expected to be of similar
significance for nuclear pion production near threshold as the
$\rho\omega$ exchange mechanism.
The $\pi\eta a_0$ vertex can be described by the Lagrangian
\begin{equation}
{\cal L}_{\pi\eta a_0}={m_{a_0}g_{\pi\eta a_0}\over m_\pi^2}
[ m_\pi^2 \vec{a}_0 \cdot \vec\pi \eta 
+ r_{\pi\eta a_0} \vec{a}_0 \cdot \partial_\mu \vec\pi \partial^\mu \eta],
\end{equation}
where $\vec{a}_0$ is the isovector $a_0(980)$ meson field and
$\eta$ the pseudoscalar $\eta$-meson field. 
Here $g_{\pi\eta a_0}$ is the non-derivative
coupling constant and $r_{\pi\eta a_0}$ is the ratio between
two-derivative and non-derivative couplings;
higher-derivative interactions could be considered as well. 
We take the decay width $\Gamma (a_0\rightarrow \pi\eta)$ of the $a_0(980)$ 
to be 175 MeV, which is the mean of the empirical
range 50-300 MeV \cite{Par}.  We emphasize that the measured value
of $\Gamma$ fixes only
one combination of the two parameters $g_{\pi\eta a_0}$ and 
$r_{\pi\eta a_0}$. The derivative term is important in the decay width 
because of the 
large three-momenta of the pion and eta, but less important near threshold
where the three-momentum of the pion vanishes. Hence 
the contribution of the $\pi\eta a_0$ vertex to $\pi^0$ production 
is very sensitive
to the parameter $r_{\pi\eta a_0}$. For example, 
with $r_{\pi\eta a_0}=0$,
one finds $|g_{\pi\eta a_0}|\simeq 3.5$ which would lead to a huge 
contribution. The  values 0 and 3.5 
are similar to those found in the linear basis of 
't Hooft's extension of the linear $\sigma$-model, which
explains the $U(1)$ anomaly \cite{tH}, and would generate a large
contribution to the pion production amplitude.
There is, however, no justification for choosing $r_{\pi\eta a_0}=0$,
as the rest of the calculation employs a pion field that couples via
derivatives  in the chiral limit. Rather, naturalness implies
$|r_{\pi\eta a_0}| \sim 1$;
e.g., in the appropriate non--linear
basis, 't Hooft's model gives $r_{\pi\eta a_0}=-2$, while
assumed meson saturation of ChPT theory parameters
in the meson sector gives \cite{EcPiRa}
$r_{\pi\eta a_0} \simeq -0.75$.
Assuming $|r_{\pi\eta a_0}|\sim 1$, the decay width is controlled
by the two-derivative term, and 
$|r_{\pi\eta a_0} g_{\pi\eta a_0}| \sim 0.21$.  \\

The $\eta N\bar N$ and $a_0 N\bar N$ coupling Lagrangians are taken to
have the usual pseudovector and scalar forms.
The coupling constant $f_{\eta NN}$ is determined as 
$f_{\eta NN}=(m_\eta/2m_N)g_{\eta NN}$, where $g_{\eta NN}=5$ \cite{MaHoEl}.
Again interpreting the $a_0$ exchange mechanism as an effective
representation of the isospin 1 scalar exchange component of the
$NN$ interaction, we take the coupling constant $g_{a_0NN}$ 
to have the value 5.9 indicated by the Bonn model \cite{MaHoEl}.
The $\eta N\bar N$ and $a_0 N\bar N$ vertex form factors are denoted 
$F_\eta(\vec{k})$ and $F_{a_0}(\vec{k})$ respectively, 
and can be obtained {}from the same reference.\\

The complete $\eta a_0$
exchange contribution to the $\pi^0$ production amplitude is then 
determined to be
\begin{equation}
T^{\eta,a_0}=-i{m_{a_0}\over m_\eta}f_{\eta NN}g_{a_0 NN}
g_{\pi\eta a_0}
{F_\eta(\vec{k})F_{a_0}(\vec{k})\over(m_\eta^2+\vec{k}^2)(m_{a_0}^2+\vec{k}^2)}
[1+ r_{\pi\eta a_0} {\omega_q^2 \over 2 m_\pi^2}]
\vec{\Sigma}\cdot \vec k,
\end{equation}
in the same notation as before.
Obviously neither the
sign nor the magnitude of this contribution is
determined by the decay width alone.
In the following, this contribution is estimated by assuming  
the magnitude of the square bracket to be 1, as suggested by naturalness,
and that the overall
sign is such that this amplitude interferes constructively with the impulse
approximation. \\

One may proceed in a similar way to incorporate exchange of
heavier mesons that can decay into $\pi^0$ and another neutral meson.
There is, however, only scant information on the coupling of such heavier
mesons to nucleons. An exception are the isovector mesons around 
1.3 GeV: the $a_1(1260)$ and the $\pi '(1300)$. Both have large decay
widths, and could be expected to decay predominantly into 
$\pi (\pi\pi)_{S-{\rm wave}}$. Arguments based on chiral
symmetry and Regge behaviour can e.g. yield a significant decay width
for $a_1(1260)\rightarrow \pi \epsilon$, where $\epsilon$ is a
broad scalar more or less degenerate with the $\rho$ \cite{We}.
An analysis that included both $a_1(1260)$ and $\pi '(1300)$
\cite{AaLo} has yielded a very small branching ratio for this process,
however, while giving $\Gamma (\pi '\rightarrow \pi \epsilon) = 345$ MeV.
To be definite,  we will concentrate on the $\pi '(1300)$, and use
this decay width as input. We consider the Lagrangian
\begin{equation}
{\cal L}_{\pi\sigma \pi '}={g_{\pi\sigma \pi '} \over f_\pi}
[ m_\pi^2 \vec{\pi}' \cdot \vec\pi \sigma 
+r_{\pi\sigma\pi '}\partial^\mu\vec{\pi}' \cdot\partial_\mu \vec\pi \sigma],
\end{equation}
where $\vec{\pi}'$ is the isovector $\pi '(1300)$ meson field and
$\sigma$ the effective scalar field. Again,
here $g_{\pi\sigma \pi '}$ is the non-derivative
coupling constant and $r_{\pi\sigma\pi '}$ is the ratio between
two-derivative and non-derivative couplings; the pion decay
constant $f_\pi$ was introduced for normalization.
And again, assuming $|r_{\pi\sigma\pi '}|\sim 1$, the decay width is 
controlled by the two-derivative term, and 
$|r_{\pi\sigma\pi '} g_{\pi\sigma\pi '}| \sim 0.7$. \\

The $\pi '(1300)$ is not included in most boson-exchange
$NN$ potentials, but 
these have typically a large $\pi NN$ form factor parameter
$\Lambda_\pi \simeq 1.3-1.5$ GeV. However, there is a considerable body of
evidence, see e.g. \cite{awt90} and references therein,
that this  form factor is actually much softer. In particular, the use
of hard pion-nucleon form factors, typical in nuclear physics, 
in calculations of deep inelastic
structure functions leads to much larger breaking of the SU(3) flavor symmetry
of the $\bar q$  sea than is observed \cite{awt83,fms89}. 
A value of 730 $\pm$ 100 MeV is preferred by those analyses. \\

The reduction of the effects of one-pion exchange
in $NN$ scattering caused by such a soft form factor
may be compensated for by
the effects of the exchange of a $\pi '(1300)$ \cite{HoTh}.
In particular, 
a form factor parameter $\Lambda_\pi =800$ MeV
can be enforced in a one-boson-exchange potential without loss of
fitting quality if the $\pi '(1300)$ is introduced simultaneously,
coupling to the 
nucleon as an ordinary pseudoscalar, with a coupling constant 
$g_{\pi 'NN}^2/4 \pi = 100$ \cite{HoTh}. In the same model, the
$\sigma$ has its usual coupling to the nucleon, with a coupling
constant $g_{\sigma NN}^2/4 \pi = 8.38$. The respective form factors
\cite{HoTh} are denoted here $F_\eta(\vec{k})$ and
$F_{a_0}(\vec{k})$. \\

In the same notation as before, the corresponding amplitude (see Fig.~1e)
is
\begin{equation}
T^{\sigma, \pi '}=-i{m_{\pi}^2\over 2 m_N f_\pi}g_{\sigma NN}g_{\pi ' NN}
g_{\pi\sigma \pi '}
{F_\sigma(\vec{k})F_{\pi '}(\vec{k})
\over(m_\sigma^2+\vec{k}^2)(m_{\pi '}^2+\vec{k}^2)}
[1+ r_{\pi\sigma \pi '} {\omega_q^2 \over 2 m_\pi^2}]
\vec{\Sigma}\cdot \vec k.
\end{equation}
Similarly to the $\eta a_0$ case, we assume in the following that
the square bracket is of $O(1)$ and indeed is taken as unity in our
calculations. We also assume that $g_{\sigma NN}g_{\pi NN} >0$ as in the linear
$\sigma$-model, and that $g_{\pi NN}g_{\pi ' NN}<0$ and 
$g_{\pi\sigma \pi '}g_{\sigma \pi\pi}<0$ as in quark models.\\

As the $N(1440)$ is the lowest ${1\over 2}^+$ nucleon resonance it
is expected to contribute to the $S$-wave pion production amplitude 
in a similar
way to the $N\bar N$ pair contributions that are implied by the
Lorentz-invariant structure of the $NN$ interaction. As the
effective pion rescattering vertices include all $\pi N$ $s$-channel
resonances, only amplitudes that arise {}from excitation of virtual
$N(1440)$ resonances by short range exchange mechanisms should be
considered as separate two-nucleon pion production amplitudes. The
most important of these are $\sigma$- (or ($\pi\pi)^{I=0}_{S-{\rm
wave}}$) and $\omega$-meson exchange (Fig.~1f).\\

The appropriate vertices for derivation of the amplitude for the 
contribution that involves excitation of intermediate $N(1440)$'s
through $\sigma$- and $\omega$-exchange are given in Ref.~\cite{CoPeRi}.
The $\sigma N\bar N$ coupling constant $g_\sigma$ is approximately given by 
the Bonn-model for the $NN$ interaction \cite{MaHoEl}. The
coupling constant $g_\sigma^*$ for the excitation of the $N(1440)$ 
resonance is estimated {}from the partial width
for the $N(1440)\rightarrow N(\pi\pi)^{I=0}_{s-{\rm wave}}$ in Ref.
\cite{CoPeRi} to be $g_\sigma^*\simeq 1.1$. The value for the $\omega
NN^*$ coupling constant, $g_\omega^*$, is highly uncertain. In Ref.
\cite{CoPeRi} it was estimated to be $g_\omega^*\simeq 1.7$ under the
assumption that $g_\omega^*/g_\omega\simeq g_\sigma^*/g_\sigma$.
The $\pi NN(1440)$ coupling is described by a pseudovector Lagrangian.
The $\pi NN^*$ pseudovector coupling constant $f_\pi^*$ is obtained as
$f_\pi^*=0.62$ {}from the empirical decay width for the
$N(1440)\rightarrow N\pi$ decay \cite{CoPeRi}.
We assume that the form factors for the $\sigma N(1440)\bar N$ 
and $\omega N(1440)\bar N$ vertices are the same as the
$\sigma N\bar N$ and $\omega N\bar N$ vertices, $F_\sigma(\vec{k})$
and $F_\omega(\vec{k})$, respectively.\\

The $N(1440)$ amplitude is then
\begin{eqnarray}
T^{R;\sigma,\omega} & = & -i {f_\pi^*\over m_N (m^*-m_N)} \omega_q
[\left(\frac{g_\sigma g_\sigma^* F_\sigma^2(\vec{k})}
            {\vec{k}^2 + m_{\sigma}^2}-
       \frac{3m_N-m^*}{m^*+m_N}
       \frac{g_\omega g_\omega^* F_\omega^2(\vec{k})}
            {\vec{k}^2 + m_{\omega}^2}\right)
                            \vec{\Sigma}\cdot\vec{P} \nonumber \\
  &   & -2 i\frac{m^*-m_N}{m^*+m_N}
    \frac{g_\omega g_\omega^* F_\omega^2(\vec{k})}{\vec{k}^2 + m_{\omega}^2}
\vec{\sigma}^{(1)}\times\vec{\sigma}^{(2)}\cdot\vec{k}].
\end{eqnarray}
The form of this operator is similar to that of the effective 
meson-exchange term that involves an intermediate $N\bar N$ pair.
The
ratio of the $N(1440)$ scalar meson-exchange part of Eq. (7) to the
scalar exchange pair contribution is  
$(g_\sigma^*/g_\sigma)(f_\pi^* f_\pi /g_A m_\pi)$ $(2m_N/(m^*-m_N))\sim 0.25$. 
This shows that the $N(1440)$ contribution enhances the scalar
meson-exchange contribution to the $S$-wave pion production amplitude
by $\sim 25\%$. 
The two vector exchange terms in Eq. (7) give contributions of different sign
to the matrix element for $pp\rightarrow pp\pi^0$. 
Their ratios to the corresponding $\omega$ pair terms are 
$-(g_\omega^*/g_\omega)(f_\pi^* f_\pi /g_A m_\pi)$ $(2m_N/(m^*-m_N))
((3m_N-m^*)/(m^*+m_N))\sim -0.15$ for the first
and 
$(g_\omega^*/g_\omega)(f_\pi^* f_\pi /g_A m_\pi)$ $(2m_N/(m^*+m_N))
\sim 0.05$ for the second. 
This partial cancellation between
the $\omega$ exchange terms that involve an intermediate
$N(1440)$ ensures that the $\omega$ exchange 
contribution to the cross section for $pp\rightarrow pp\pi^0$
near threshold will not be very different {}from the one obtained in
Ref. \cite{LeRi}.\\

All the above contributions, together with the pair terms of
Ref. \cite{LeRi}, correspond 
to contact $\pi NNNN$ interactions in ChPT.
According to the modified chiral power counting developed 
in Ref. \cite{CoFrMiKo}, the pair and $N(1440)$ terms contribute
to the leading order of such contact terms, while the exchange mechanisms
contribute to higher orders, i.e. they are down by one power of $m_\pi$.
In that reference,
the cross section for the reaction $pp\rightarrow pp\pi^0$ 
was calculated on the basis of a reaction amplitude formed
primarily of (1) the single nucleon pion production operator, (2) the
pion exchange contribution as constrained by chiral
perturbation theory, and (3) the short range contributions associated
with intermediate $N\bar N$ pair terms.
It was shown that the resulting cross section is too small by 
a factor $5-10$ 
because of the remarkable cancellation between the single nucleon and pion 
rescattering contributions. This cancellation depends crucially
on parameters determined by $\pi N$ scattering. \\

We shall here be mainly concerned with the modifications caused by 
(i) the additional short-range contributions discussed above
and (ii) changes in the numerical values of the parameters
that affect the pion rescattering contribution. \\

The numerical values of the different contributions
of the graphs of Fig.~1 are shown in Fig.~2. 
These are computed using the Reid93 potential \cite{Fr}
to provide the
initial $^3P_0$ and final $^1S_0$ wave functions. The quantities $J$ refer
to the amplitudes after some common kinematic factors and complex phases 
are removed; see Ref. \cite{CoFrMiKo} for a precise definition.
The label impulse refers to the contribution of Fig.~1a.
There are two pion rescattering curves. They both include the same
$\Delta$ excitation contribution (Fig. 1c) and a contribution (Fig. 1b)
proportional to 
$$\left[4c_1+{\delta m_N\over 2 m_\pi^2}
-\left(c_2+c_3-{g_A^2\over8m_N}\right)\right]= {C\over 2\;m_N},$$
where $c_i$ are input
$\pi N$ parameters, and $\delta m_N \sim 3$ MeV 
is the strong interaction contribution
to the neutron-proton mass difference. 
The two pion rescattering curves differ in the $c_i$'s and $C$.
The one labelled cl (for chiral loops) uses the  value of 
$C =-2.31$ obtained in Ref. \cite{CoFrMiKo}
and corresponds to the $O(q^3)$ parameters
$c_i$ as given in Ref. \cite{BeKaMe1}.
It is obvious
that the contribution of the impulse term is almost entirely
cancelled by this pion rescattering term.
This is why the corresponding cross section in 
Ref. \cite{CoFrMiKo} was very close
to zero without including the effects of the nucleon-pair terms associated 
with $\sigma$ and $\omega$ exchange. 
The other rescattering curve (labelled st for sub-threshold expansion)
corresponds to the $O(q^2)$ parameters
$c_i$ as given in Ref. \cite{BeKaMe2}, which give a much smaller $C= -0.29$.
(The subthreshold expansion makes use of an analytic description of the
$\pi N$ with numerical values determined by data
\cite{HO}.)
As a consequence the cancellation is less complete in this case.
Both of these curves were obtained with a soft pion form factor;
employing a hard form factor increases the magnitude of these results  
by less than 10\%. 
This is because of the effects of $NN$ repulsion
in the final state $^1S_0$ $pp$ wave function and the centrifugal 
barrier in the
initial $^3P_0$ wave function. The integrand for the pion rescattering
contribution peaks at about 1.3 fm.\\

The other curves in Fig.~2 display the shorter-range effects
(Fig.~1d,e,f).
The effect of the new $\rho\omega\pi^0$ has 
the same sign as the impulse term 
and is comparable to
the change in the rescattering term; it is also of similar size
as the $\sigma$ pair term.
The new $\eta a_0\pi^0$ 
and $\sigma \pi '\pi^0$ meson exchange terms are smaller under the
(conservative) assumptions of this paper. 
The curve labelled $\omega + \sigma$ 
includes the effects of the pair diagrams. The effects of 
the excitation of the $N(1440)$ are also shown. 
Including the $N(1440)$ enhances the sigma exchange term by about 25\%,
but the omega exchange term hardly at all because of the
cancellation noted above. Therefore the effect of the
$N(1440)$ is small. The results in 
Fig.~2, do have a wide uncertainty margin because 
of the uncertain meson couplings.
In particular, using a value of $r_{\pi\sigma\pi'}$ much smaller than
unity would increase the value of the related contribution by a big factor.
Likewise, a larger value for $g_\sigma^*$ \cite{Os} 
would also increase the cross section.\\

We also note that the magnitude of these
shorter-than-pion-range terms is consistent with expectations {}from 
the modified chiral power counting of Ref. \cite{CoFrMiKo}. 
Although the convergence of ChPT here 
is not that impressive, there is no evidence of failure
either.\\

The cross sections, computed using all of the mechanisms of Fig.~1 as well 
as
the very small recoil term of Ref. \cite{CoFrMiKo} are shown in Fig.~3,
together with experimental data {}from Refs. \cite{Me,Mey,Bo}.
The full curves represent the complete result for the
Reid93 potential \cite{Fr}, the lower curve using the $O(q^3)$ rescattering
parameters, the higher the $O(q^2)$ parameters.
Also shown are the dashed curves with the corresponding results for the
Argonne V18 potential \cite{WiStSc}.
For a given potential, the two different set of rescattering parameters
give an estimate of the uncertainty generated by the rescattering
mechanism. For a given set of rescattering parameters, the two potentials
give an estimate of the uncertainty generated by the nuclear short-range 
dynamics. The overall theoretical uncertainty {}from these sources
is quite large; it arises because the cross section is basically very small. 
The inputs (both $pp$ wave functions and various $\pi N$
scattering coefficients)
are not known well enough to provide a more accurate calculation. There is also
additional uncertainty in the input coefficients of the meson-exchange
Lagrangians of Eqs. (3) and (5).
Furthermore, two-pion exchange diagrams can be expected to contribute
appreciably, but have so far not been calculated.
Within this large inherent theoretical uncertainty, meson exchange
can explain the observed cross section.  \\

Further development of ChPT should lead to a reduction of the  
uncertainties associated with the
$\pi N$ scattering coefficients, $pp$ wave functions,
and the mesonic  coupling constants.
A calculation to order ($q^3$) 
and comparison with the sub-threshold expansion 
of Ref. \cite{HO} should lead to better 
description of $\pi N$ scattering  data. If the chiral potential of 
Ord\'o\~nez et al \cite{bira} were developed to
yield a good description of
$NN$ data up to the energies relevant here
a consistent treatment of the amplitude for
$\pi^0$ production and the nuclear wave functions based on
the same general chiral Lagrangian could be achieved.
Finally an improved treatment 
of the meson-exchange diagrams is possible by a
chiral calculation of the relevant 
terms as higher order
loop diagrams. \\

The principal conclusion is that the influence of the $\rho\omega\pi^0$ 
-- and to a lesser extent the other shorter-than-pion-range mechanisms 
considered here --
is considerable.
This idea can be used to understand independent experiments. In particular, 
the PCAC relation between the axial charge operator
of the nuclear system and the S-wave soft-pion production
amplitude \cite{LeRi} and the present results imply that the axial
charge operator should have a significant $\rho\omega$ 
axial exchange current term. As this will
contribute to the nuclear enhancement of the effective
axial charge of the nucleon \cite{KiRiTs}, this suggests a 
possibility for further reducing the remaining
uncertainties in the magnitude of the short range mechanisms
by data on first forbidden nuclear $\beta$-transitions 
\cite{Wa,War}.\\

This work is partially supported by the USDOE. G.A.M. is grateful for the 
hospitality of T.E.O. Ericson and  the Nobel Foundation during a visit to
Uppsala University where part of this work was completed.\\

\newpage

{\bf Figure captions}
\vspace{1cm}

Fig. 1 \hspace{0.5cm} 
Contributions to the amplitude for
$pp\rightarrow pp\pi^0$. (a) Impulse approximation, (b) pion rescattering,
(c) pion rescattering through an intermediate $\Delta$, (d) pair graphs,
(e) meson exchange contributions, (f) excitation of an intermediate 
$N(1440)$ resonance.\\

Fig. 2 \hspace{0.5cm} 
Individual contributions to the 
amplitude for $pp\rightarrow pp\pi^0$ as a function of $p'$, the relative
momentum between the two final protons for laboratory kinetic energy of 
300MeV. The Reid93 potential is used.\\

Fig. 3 \hspace{0.5cm} 
Computed cross sections for two different $NN$ potential models.
All of the terms of Fig. 1 are included as well as that of
the recoil term of Ref. [2].\\

\newpage

\begin{figure}
\centerline{\epsffile{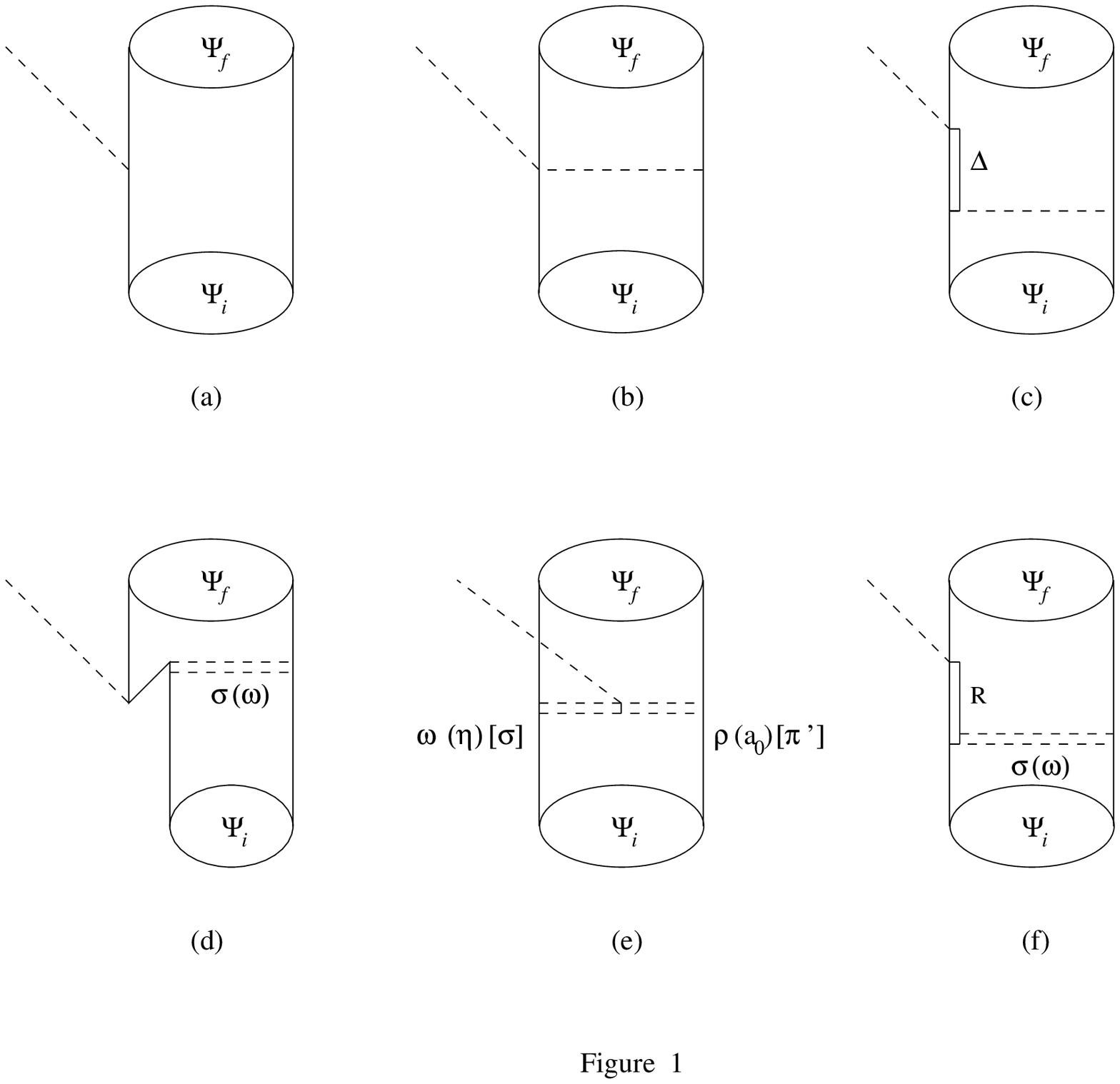}}
\end{figure}

\newpage

\begin{figure}
\vspace*{-4cm}
\centerline{\epsffile{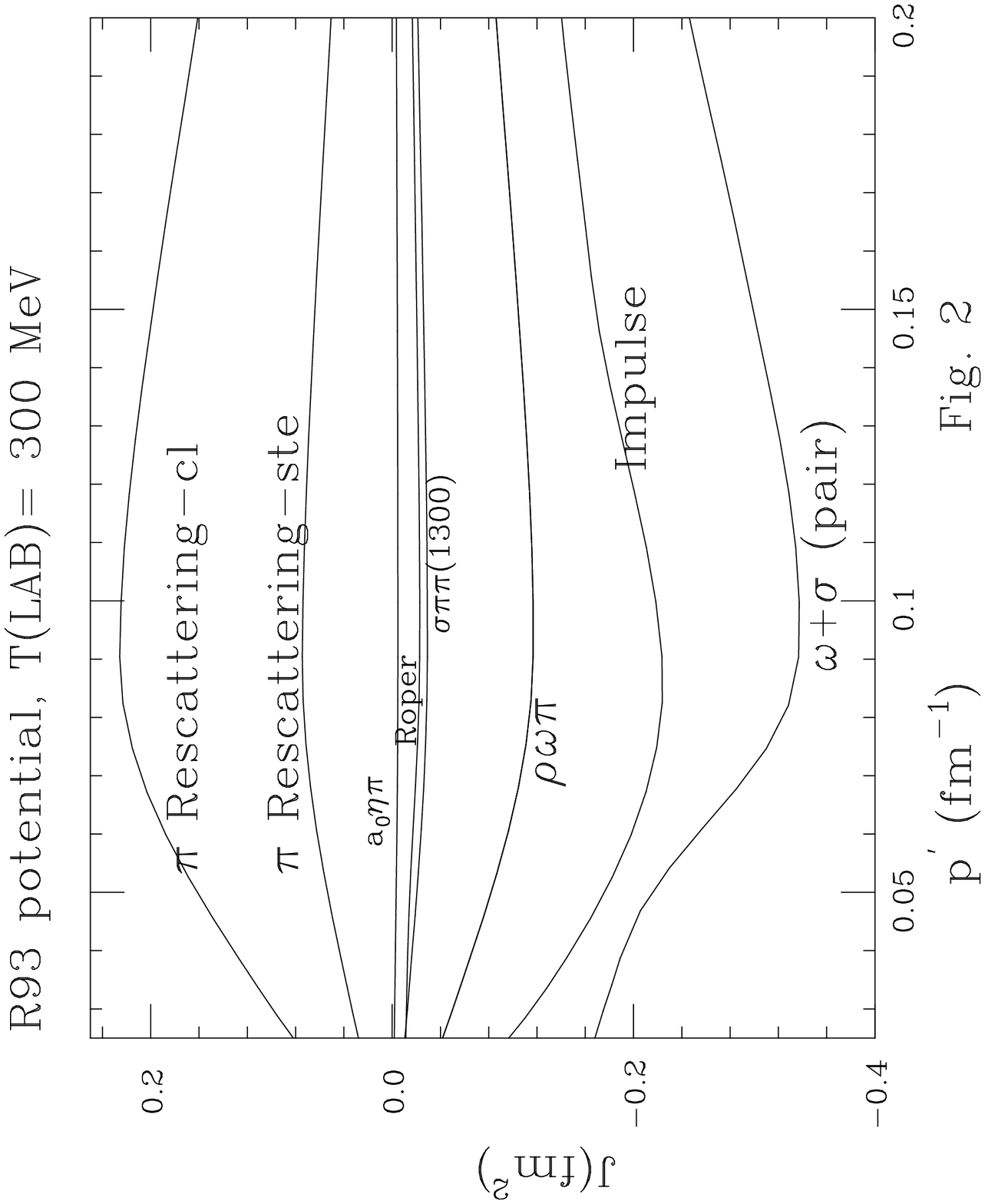}}
\end{figure}

\newpage

\begin{figure}
\vspace*{-4cm}
\centerline{\epsffile{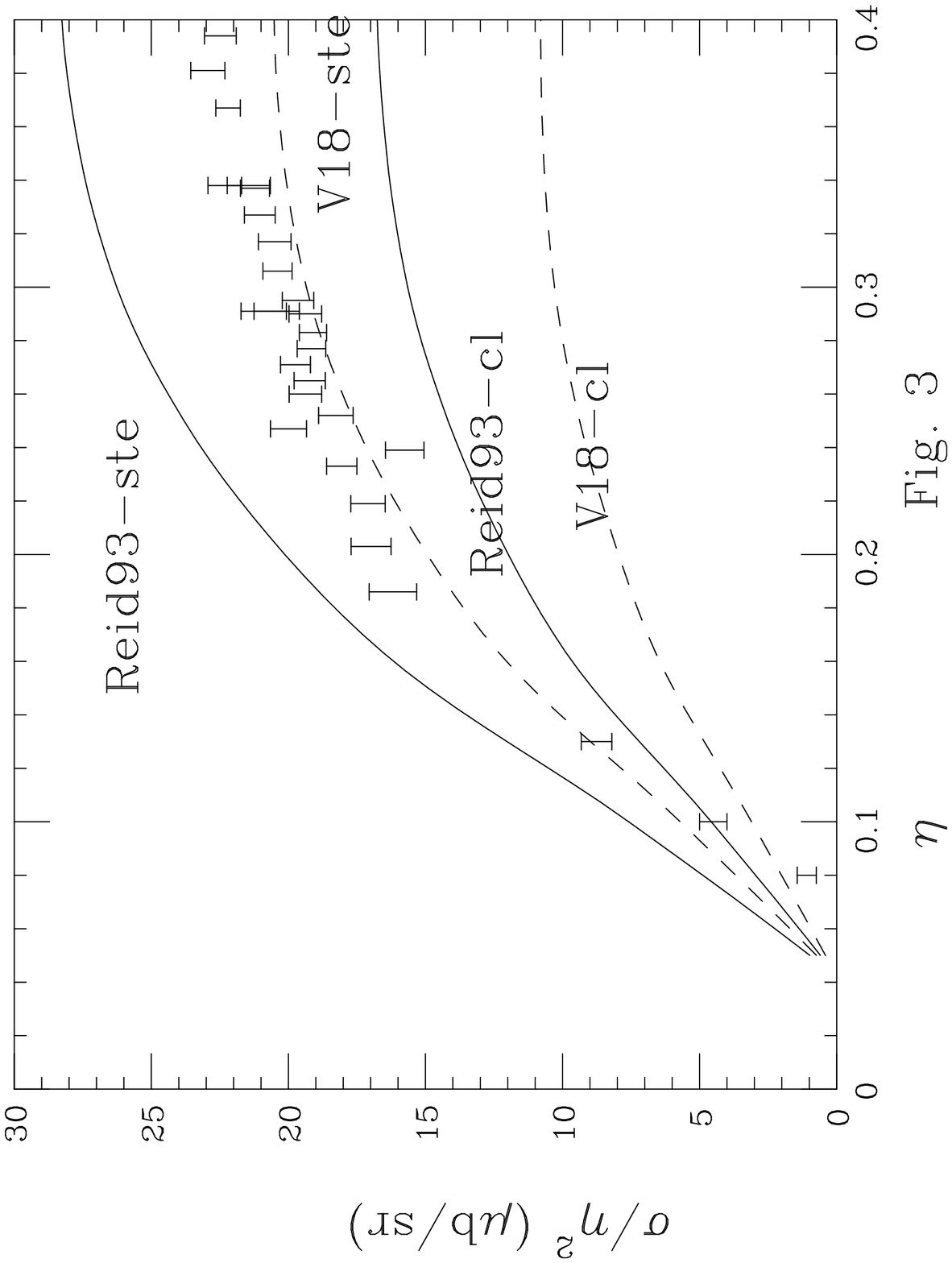}}
\end{figure}

\end{document}